\documentclass[a4paper,10pt]{article}
\usepackage[margin=2cm]{geometry}
\usepackage[english]{babel}
\usepackage[latin1]{inputenc}

\newcommand{\pder}[2]{\frac{\displaystyle\partial#1}{\displaystyle\partial#2}}
\newcommand{\dfrac}[2]{\frac{\displaystyle#1}{\displaystyle#2}}

\title{Hybrid modeling of plasmas}
\date{January 12, 2010}
\author{Mats Holmstr{\"o}m\thanks{Swedish Institute of Space Physics, PO~Box~812, SE-98128~Kiruna, Sweden, 
\texttt{matsh@irf.se}}}

\begin{document}

\maketitle

\section{Introduction}
Space plasmas are often modeled as a magnetohydrodynamic (MHD) fluid.  However, many observed phenomena cannot be captured by fluid models, e.g., non-Maxwellian velocity distributions and finite gyro radius effects.  Therefore kinetic models are used, where also the velocity space is resolved.  This leads to a six-dimensional problem, making the computational demands of velocity space grids prohibitive.  
Particle in cell (PIC) methods discretize velocity space by representing the charge distribution as discrete particles, and the electromagnetic fields are stored on a spatial grid.  For the study of global problems in space physics, such as the interaction of a planet with the solar wind, it is difficult to resolve the electron spatial and temporal scales.  Often a hybrid model is then used, where ions are represented as particles, and electrons are modeled as a fluid.  Then the ion motions govern the spatial and temporal scales of the model. 
Here we present the mathematical and numerical details of a general hybrid model for plasmas.  All grid quantities are stored at cell centers on the grid.  
The most common discretization of the fields in PIC solvers is to have the 
electric and magnetic fields staggered, introduced by Yee \cite{Yee}. 
This automatically ensures that $\nabla\cdot\mathbf{B}=0$, down to round-off errors.  Here we instead present a cell centered discretization of the magnetic field.  That the standard cell centered second order stencil for $\nabla\times\mathbf{E}$ in Faraday's law will preserve $\nabla\cdot\mathbf{B}=0$ was noted by \cite{Toth}.  The advantage of a cell centered discretization is ease of implementation, and the possibility to use available solvers that only provide for cell centered variables.  We also show that the proposed method has very good energy conservation for a simple test problem in one-, two-, and three dimensions, when compared to a commonly used algorithm.

\section{Definitions}
We have $N_I$ ions at positions $\mathbf{r}_i(t)$~[m] with velocities 
$\mathbf{v}_i(t)$~[m/s], mass $m_i$~[kg] and charge $q_i$~[C], 
$i=1,\ldots,N_I$. 
By spatial averaging\footnote{Usually, charge and current densities are 
deposited on a grid, using shape functions \cite{hoc89}.}, 
we can define the 
charge density $\rho_I(\mathbf{r},t)$~[Cm$^{-3}$] of the ions, 
their average velocity $\mathbf{u}_I(\mathbf{r},t)$~[m/s], and 
the corresponding current density 
$\mathbf{J}_I(\mathbf{r},t)=\rho_I\mathbf{u}_I$~[Cm$^{-2}$s$^{-1}$]. 
Electrons are modelled as a fluid with charge density 
$\rho_e(\mathbf{r},t)$, average velocity $\mathbf{u}_e(\mathbf{r},t)$, 
and current density $\mathbf{J}_e(\mathbf{r},t)=\rho_e\mathbf{u}_e$. 
The electron number density is $n_e=-\rho_e/e$, where $e$ is 
the elementary charge. 
If we assume that the electrons are an ideal gas, 
then $p_e=n_ekT_e$, so the pressure is directly related to temperature 
($k$ is Boltzmann's constant).

The trajectories of the ions are computed from the Lorentz force, 
\[
  \dfrac{d\mathbf{r}_i}{dt} = \mathbf{v}_i, \quad
  \dfrac{d\mathbf{v}_i}{dt} = \dfrac{q_i}{m_i} \left( 
    \mathbf{E}+\mathbf{v}_i\times\mathbf{B} \right), \quad
    i=1,\ldots,N_I
\]
where $\mathbf{E}=\mathbf{E}(\mathbf{r},t)$ is the electric field, 
and $\mathbf{B}=\mathbf{B}(\mathbf{r},t)$ 
is the magnetic field.\footnote{\cite{bag02} modifies the electric 
field in the Lorentz force by a term proportional to $\cal{C}$ and 
$\nabla\times\mathbf{B}$ to preserve momentum.}

\subsection{Hybrid approximations}
A brief overview of hybrid codes can be found in~\cite{win01}. 
A more complete survey can be found in \cite{pri00}. 
Most hybrid solvers for global simulations have the following 
assumptions in common. 
\begin{enumerate}
  \item Quasi-neutrality, $\rho_I+\rho_e = 0$, so that 
        given the ion charge density, the electron charge density 
        is specified by $\rho_e=-\rho_I$. 
  \item Ampere's law whithout the transverse displacement current 
        (also called the Darwin approximation, or the nonradiative limit) 
        provides the total current, given $\mathbf{B}$, by 
        \[
          \mathbf{J} = \mu_0^{-1}\nabla\times\mathbf{B}, 
        \]
        where $\mu_0=4\pi\cdot10^{-7}$~[Hm$^{-1}$] is the 
        magnetic constant ($\epsilon_0\mu_0c^2=1$), 
        and from the total current we get the electron current, 
        $\mathbf{J}_e = \mathbf{J}-\mathbf{J}_I$, and thus the electron 
        velocity, since the quasi-neutrality implies that 
        $\mathbf{u}_e = \mathbf{J}_e/\rho_e = 
         (\mathbf{J}_I-\mathbf{J})/\rho_I$. 
  \item Massless electrons, $m_e = 0$, lead to the electron momentum equation
    \[
      n_em_e\dfrac{d\mathbf{u}_e}{dt} = \mathbf{0} = \rho_e\mathbf{E} 
      +\mathbf{J}_e\times\mathbf{B} -\nabla p_e + \cal{C}
    \]
    where the force terms $\cal{C}$ can be due to collisions, such as 
    electron-ion collisions, electron-neutral \cite{ter02} collisions, 
    or anomalous, 
    i.e.\ representing electron-wave interactions \cite{bag02}. 
    In our numerical experiments we have assumed that $\cal{C}=0$. 
    This provides an equation of state (Ohm's law) for the electric field
    \[
      \mathbf{E} = \frac{1}{\rho_I} \left[ (\mathbf{J}-\mathbf{J}_I)  
                   \times \mathbf{B} -\nabla p_e + \cal{C} \right], 
    \]
    with $\mathbf{J}$ from Ampere's law.  
    So the electric field is not an unknown.  Whenever it is needed, 
    it can be computed. 
  \item Faraday's law is used to advance the magnetic field in time, 
        \[
          \pder{\mathbf{B}}{t} = -\nabla\times\mathbf{E}. 
        \]
  \item The electron pressure is isotropic ($p_e$ is a scalar, not a tensor). 
\end{enumerate}
For the electrons, the remaining degree of freedom is the pressure, 
$p_e$.
Note that $p_e$ only affects the ion motions through the electric field.  
The evolution of the magnetic field is not affected since we have 
$\nabla\times\nabla p_e=0$ in Faraday's law. 
There are several ways to handle the electron pressure~\cite[p.\ 8790]{win86}, 
\begin{enumerate}
\item Assume $p_e$ is constant, or zero \cite{kal03}. 
\item Assume $p_e$ is adiabatic (small collision frequency).  
      Then the electron pressure is related to the electron 
      charge density by $p_e\propto|\rho_e|^\gamma$,  
      where $\gamma$ is the adiabatic index.  
      Commonly used values are $\gamma=5/3$ \cite{bag02,lip02}, 
      and $\gamma=2$ \cite{rou08,bos04}. 
\item Solve the massless fluid energy equation~\cite{ric94,lip02}, 
      \[
        \pder{p_e}{t} + \mathbf{u}_e\cdot\nabla p_e + \gamma p_e 
        \nabla\cdot \mathbf{u}_e = 
        \left(\gamma-1\right) \eta |\mathbf{J}|^2, 
      \] 
\end{enumerate}
Here we assume that $p_e$ is adiabatic.  Then the relative change in 
electron pressure is related to the relative change in electron density by 
\[
  \frac{p_e}{p_{e0}} = \left( \frac{n_e}{n_{e0}} \right)^{\gamma} , 
\]
where the zero subscript denote reference values. 
From charge neutrality and $p_e=n_ekT_e$ we have that 
\[
  p_e = A \rho_I^{\gamma} 
    \mbox{ with } A=\frac{k}{e}\rho_I^{1-\gamma}T_e 
\]
a constant that is evaluated using reference values of $\rho_I$ and $T_e$, 
e.g., solar wind values. 
Note that $\gamma=1$ corresponds to assuming that $T_e$ is constant, and 
$\gamma=0$ gives a constant $p_e$. 

\subsubsection{Hybrid equations}
If we store the magnetic field on a discrete grid $\mathbf{B}_j$, 
the unknowns are $\mathbf{r}_i$, $\mathbf{v}_i$, and $\mathbf{B}_j$ 
(supplemented by $p_e$ on a grid, if we include the electron energy equation). 
The time advance of the unknowns can then be written as the ODE 
\begin{equation}
\dfrac{d}{dt}
\left( \begin{array}{c}
           \mathbf{r}_i \\ \mathbf{v}_i \\ \mathbf{B}_j \\ 
         \end{array} \right)
=
\left( \begin{array}{c}
           \mathbf{v}_i \\ 
           \dfrac{q_i}{m_i} 
           \left( \mathbf{E}+\mathbf{v}_i\times\mathbf{B} \right) \\ 
           -\nabla_j\times\mathbf{E} \\ 
         \end{array} \right)
\end{equation}
where $\nabla_j\times$ is a discrete rotation operator, and the electric 
field is 
\[
  \mathbf{E}_j = \dfrac{1}{\rho_I} \left( -\mathbf{J}_I\times\mathbf{B}_j 
  +\mu_0^{-1}\left(\nabla_j\times\mathbf{B}_j\right) \times \mathbf{B}_j 
  \right) - \nabla p_e  + \cal{C}. 
\]

\section{Discretisation}
An overview of different discretizations of the above equations 
can be found in \cite[Appendix~A]{kar04}. 
\cite[Section~3.1]{bag02} provides a consise description of the CAM-CL 
algorithm introduced by \cite{mat94}. 
All our grid variables will be cell centered: 
$\mathbf{B}_j$, $\mathbf{J}_j$, and $\mathbf{\rho}_j$
(here $\mathbf{J}_j$ and $\mathbf{\rho}_j$ are the ionic current and the ionic 
charge density at cell centers --- from now we omit the subscript $I$ for 
simplicity).  We follow the Current Advance Method and Cyclic Leapfrog 
(CAM-CL) algorithm~\cite{mat94}, but omit the CAM part.  
The Current Advance Method is used to avoid multiple iterations over 
the particles, but does not conserve energy well, as we will see for 
a test problem. 

For the particles, we have to solve for $\mathbf{r}_i$ and $\mathbf{v}_i$; and 
for the grid cells we have to solve for $\mathbf{B}_j$, $\mathbf{J}_j$ and 
$\mathbf{\rho}_j$.  We denote time level $t=n\Delta t$ by superscript $n$. 
Given $\mathbf{B}_j^{n-1/2}$, $\mathbf{r}_i^{n-1/2}$ and $\mathbf{v}_i^n$, 
we do the following steps. 
\[
  \mathbf{r}_i^{n+1/2} \leftarrow \mathbf{r}_i^{n-1/2} 
    + \Delta t \mathbf{v}_i^n, 
\]
\[
  \mathbf{r}_i^{n} \leftarrow \frac{1}{2}\left(\mathbf{r}_i^{n+1/2} + 
                              \mathbf{r}_i^{n-1/2} \right). 
\]
At $\mathbf{r}_i^n$, deposit particle charges and currents 
\[
  \rho_i \rightarrow \rho_j^{n}, \qquad
  \rho_i\mathbf{v}_i^{n} \rightarrow \mathbf{J}^{n}_j,
\]
\begin{equation} \label{eq:cl}
  \mathbf{B}_j^{n+1/2} \leftarrow 
       \mathbf{B}_j^{n-1/2}, \rho_j^n, \mathbf{J}^n_j
       \qquad\mbox{according to CL.} 
\end{equation}
At $\mathbf{r}_i^{n+1/2}$, deposit particle charge 
\[
  \rho_i \rightarrow \rho_j^{n+1/2}. \qquad
\]
Estimate electric field at $n+1/2$ using the currents at $n$ 
\[
  \mathbf{E}_j^{*} \leftarrow 
       \mathbf{B}_j^{n+1/2}, \rho_j^{n+1/2}, \mathbf{J}^{n}_j, p_e,
\]
\[
  \mathbf{v}_i^{n+1/2} \leftarrow \mathbf{v}_i^{n} + \frac{\Delta t}{2}
    \frac{q_i}{m_i} \left( \mathbf{E}_j^{*} 
    + \mathbf{v}_i^n \times \mathbf{B}_j^{n+1/2}  \right).
\]
At $\mathbf{r}_i^{n+1/2}$, deposit particle current 
\[
  \rho_i\mathbf{v}_i^{n+1/2} \rightarrow \mathbf{J}^{n+1/2}_j,
\]
\[
  \mathbf{E}_j^{n+1/2} \leftarrow 
       \mathbf{B}_j^{n+1/2}, \rho_j^{n+1/2}, \mathbf{J}^{n+1/2}_j, p_e
\]
\[
  \mathbf{v}_i^{n+1} \leftarrow \mathbf{v}_i^{n} + \Delta t 
    \frac{q_i}{m_i} \left( \mathbf{E}_j^{n+1/2} 
    + \mathbf{v}_i^{n+1/2} \times \mathbf{B}_j^{n+1/2}  \right).
\]
Now we have $\mathbf{B}_j^{n+1/2}$, $\mathbf{r}_i^{n+1/2}$ and 
$\mathbf{v}_i^{n+1}$.  Set $n \leftarrow n+1$ and start over again. 

For each particle we need a temporary vector. 
First $\mathbf{r}_i^{n+1/2}$ is temporarily saved during the deposit 
at $\mathbf{r}_i^{n}$.  Then $\mathbf{v}_i^{n}$ is temporarily 
saved until the final velocity update. We also need to store the current 
corresponding to each particle, $\rho_i\mathbf{v}_i^{*}$, 
in preparation of the deposit operations. 

The update of the magnetic field in~(\ref{eq:cl}) using cyclic leapfrog (CL) 
is done in $m$ sub-time steps of length $h=\Delta t/m$. 
With the notation 
$\mathbf{B}_j^{p}\equiv\mathbf{B}_j\left( (n+1/2)\Delta t+ph \right)$
we have the iteration
\[  \renewcommand{\arraystretch}{1.5}
\left\{
  \begin{array}{l}
    \mathbf{B}_j^{1} \leftarrow \mathbf{B}_j^{0} 
        - h\nabla\times\mathbf{E}_j^{0}, \\
    \mathbf{B}_j^{p+1} \leftarrow \mathbf{B}_j^{p-1} 
        - 2h\nabla\times\mathbf{E}_j^{p}, \qquad p=1,2,\ldots,m-1, \\
    \tilde{\mathbf{B}}_j^{m} \leftarrow \mathbf{B}_j^{m-1} 
        - h\nabla\times\mathbf{E}_j^{m}, \quad\mbox{and} \\
    \mathbf{B}_j^{n+1/2} \leftarrow \frac{1}{2}\left(\mathbf{B}_j^{m} 
        + \tilde{\mathbf{B}}_j^{m} \right). \\
  \end{array}
\right.
\]
Since the magnetic field is leapfrogged in time, 
we need one temporary grid cell vector.

\subsection{Non-periodic boundary conditions}
To be able to model the interaction of objects with the solar wind, 
non-periodic boundary conditions in the $x$-direction have been 
implemented.  At $x_{\mbox{\tiny min}}$ we have an inflow boundary, and at 
$x_{\mbox{\tiny max}}$ an outflow boundary.  The other boundaries are still 
periodic. The computation of $\nabla\times\mathbf{E}$ in the interior of 
the simulation domain requires $\mathbf{E}$ 
in one extra layer of cells in the $x$-directions.  Also, computing 
$\mathbf{E}$ in the interior of the simulation domain involves 
$\nabla\times\mathbf{B}$, thus also requiring $\mathbf{B}$ in one outer 
layer of cells. 
At the inflow boundary we specify solar wind values of 
$\mathbf{B}$ and $\mathbf{E}=-\mathbf{u}_I\times\mathbf{B}$.  
At the outflow boundary we extrapolate $\mathbf{E}$ and $\mathbf{B}$ from 
the interior of the simulation domain to one external cell layer 
(a simple copy of the values from the upstream cells).

\subsection{Spatial and temporal scales}
If we want solutions of the discrete equations to be accurate 
approximations of the solutions to the continuous equations, 
a necessary condition is that the discretisation resolves all 
relevant spatial and temporal scales. 
The smallest spatial scale for the hybrid equations is 
the ion inertial length (the ion skin depth) $\delta_i=c/\omega_{pi}$, 
where $c$ is the speed of light and $\omega_{pi}$ is the ion plasma frequency, 
$\omega_{pi}^2=n_iq_i^2/(\epsilon_0 m_i)$, $n_i$ the ion number density, 
$q_i$ the ion charge, $m_i$ the ion mass, and 
$\epsilon_0\approx8.854\cdot10^{-12}$~[Fm$^{-1}$] the vacuum permittivity. 
The ion inertial length is associated with the $\mathbf{J}\times\mathbf{B}$
term in Ohm's law (the Hall term) that describes whistler dynamics. 
The fastest temporal scale is also associated with whistler dynamics. 
The whistler wave spectrum is cutoff at the electron cyclotron frequency, 
but due to the assumption of massless electrons it is unbounded for the 
hybrid equations, 
and the frequency scales like $\omega/\Omega_i=(kc/\omega_{pi})$ for 
large $k$ \cite{pri00}. 
Here $\Omega_i=q_iB/m_i$ is the ion gyrofrequency. 
This gives the CFL constraint
\[
  \Delta t < \frac{ \Omega_i^{-1} }{\sqrt{n}\pi}
    \left( \frac{\Delta x}{\delta_i} \right)^2 
\]
where $n$ is the spatial dimension.

\section{A quiet plasma test problem}
A uniform, or quiet, plasma is a first test of any simulation code. 
The solution should only show small statistical fluctuations, 
and energy should be preserved for long simulation times. 
Matthews~\cite{mat94} describes one- and two-dimensional quiet plasma runs, 
and Brecht~\cite{brecht06} present three-dimensional results. 

The number of cells used here is 16, $64^2$, and $32^3$.  
All boundary conditions are periodic. 
Ion and electron temperatures are given by, $\beta_i=1$, and $\beta_e=0$. 
Brecht~\cite{brecht06} uses a transport equation for the electron temperature. 
The number of magnetic field sub cycles is 4 in \cite{mat94}, 
3 here, and \cite{brecht06} does not use sub cycling. 

Total energy, the sum of the energy stored in the electric and magnetic 
fields and the kinetic energy of the particles, should be conserved. 
\begin{table}
\caption{Energy errors (total energy) for quiet plasma runs at times $T$. 
Numbers in parentheses indicate that the parameter was not stated in the 
reference. 
}
\begin{center}
\begin{tabular}{l|c|c|c|c|c|c|c}   \label{tab:quiet}
Reference & dim. & particles & $\Delta x$ & $\Delta t$ & T & error & error \\
     &      & per cell  & $\delta_i$ & $\Omega_i^{-1}$ & $\Omega_i^{-1}$ & Ref.
  & Here  \\ \hline
\cite{mat94}    & 1 & 16 & 0.5 & 0.1 & 100 & 9\%   & 0.9\% \\ 
                 &   &    &     &     & 300 & 47\%  &   3\% \\
                 & 2 & 32 & 0.5 & 0.1 & 100 & 2.6\% & 0.9\% \\ 
                 &   &    &     &     & 300 & 14\%  &   3\% \\
\cite{brecht06} & 3 &  4 & (1.54) & 0.0056 & 112 & $<$1\% & 0.25\% \\ 
\end{tabular}
\end{center}
\end{table}
In Table~\ref{tab:quiet} we compare the relative errors in total energy with 
the published values in one-, two-, and three dimensions.

\section{Conclusions}
The hybrid method stores the magnetic field on a grid. 
Here we have presented a cell centered algorithm as an alternative 
to the staggered grid commonly used.  The cell centered method 
preserves $\nabla\cdot\mathbf{B}=0$ down to round-off errors.  
In Table~\ref{tab:quiet} it is evident that the proposed method 
conserves energy well when compared to the commonly used 
CAM-CL method~\cite{mat94}. 
That the CAM-CL method does not conserve energy well has been noted 
before~\cite{brecht06,KraussVarban05}. 

\section*{Acknowledgements}
This research was conducted using the resources of the 
High Performance Computing Center North (HPC2N), Ume\aa\ University, Sweden, 
and the Center for Scientific and Technical Computing (LUNARC), 
Lund University, Sweden. 
The software used in this work was in part developed by the 
DOE-supported ASC / Alliance Center for Astrophysical 
Thermonuclear Flashes at the University of Chicago.


\end{document}